
\input phyzzx.tex
\input tables.tex

\def\chitil{\widetilde\chi}
\def\cnone{\chitil^0_1}

\def\ptr{p_T}
\def\mtwo{M_{jj}}
\def\mthree{M_{bjj}}
\def\mmissl{M_{miss-\ell}}

\def\delmw{\Delta\mw}
\def\delmt{\Delta\mt}
\def\gev{~{\rm GeV}}

\def\fbi{~{\rm fb}^{-1}}


\def\prdj#1{{\it Phys. Rev.} {\bf D{#1}}}
\def\npbj#1{{\it Nucl. Phys.} {\bf B{#1}}}

\def\plbj#1{{\it Phys. Lett.} {\bf B{#1}}}
\def\zpcj#1{{\it Z. Phys.} {\bf C{#1}}}

\def\mt{m_t}

\def\etmiss{E_T^{\,miss}}
\def\etell{E_T^{\,\ell}}
\def\ptmiss{\vec p_T^{\,\,miss}}
\def\ptell{\vec p_T^{\,\,\ell}}
\def\rta{\rightarrow}

\def\hn{h}

\def\mhn{m_\hn}

\def\nsd{N_{SD}}

\def\mw{m_W}

\def\anti{\overline}

\def\ifmath#1{\relax\ifmmode #1\else $#1$\fi}

\def\3quarter{{\textstyle{3 \over 4}}}

\def\tth{t\anti t \hn}
\def\ttz{t\anti t Z}
\def\ttglnu{t\anti t g(\rta\ell\nu\ell\nu)}
\def\ttbdecay{t\anti t g(b\rta\ell\nu)}

\input phyzzx
\Pubnum={$\caps UCD-93-28$\cr}
\date{August, 1993\cr revised: December, 1993}

\titlepage
\vskip 0.75in
\baselineskip 0pt
\hsize=6.5in
\vsize=8.5in
\centerline{{\bf Detecting an Invisibly Decaying Higgs Boson at
a Hadron Supercollider}}
\vskip .075in
\centerline{ J.F. Gunion}
\vskip .075in
\centerline{\it Davis Institute for High Energy Physics,
Dept. of Physics, U.C. Davis, Davis, CA 95616}

\vskip .075in
\centerline{\bf Abstract}
\vskip .075in
\centerline{\Tenpoint\baselineskip=12pt
\vbox{\hsize=12.4cm
\noindent We demonstrate that an invisibly decaying Higgs boson
with Standard Model coupling strength to $t\anti t$ can be
detected at the LHC for masses $\lsim 250\gev$.
}}

\vskip .15in
\noindent{\bf 1. Introduction and Procedure}
\vskip .075in

Perhaps the most fundamental mission of future high energy
hadron supercolliders such as the LHC is the detection
of Higgs boson(s). While many production/decay modes have been
studied in the past, an invisibly decaying Higgs boson ($\hn$) has
not been thoroughly studied. Dominance of the decays of a Higgs boson
by invisible channels is possible, in particular, in supersymmetric models.
If R-parity is conserved, decays to $\cnone\cnone$, where $\cnone$
is the lightest supersymmetric particle, can be dominant.
\Ref\susyinvisible{See, for instance: K. Greist and H.E. Haber,
\prdj{37} (1988) 719; A. Djouadi, J. Kalinowski,
and P.M. Zerwas, \zpcj{57} (1993) 569; and J.L. Lopez, D.V. Nanopoulos,
H. Pois, X. Wang, and A. Zichichi, preprint CERN-PPE-93-16 (1993).}
Indeed, in the minimal supersymmetric model $\cnone\cnone$
dominance is possible for both the lightest CP-even Higgs boson
and the CP-odd Higgs boson.
In supersymmetric models with spontaneously
broken R-parity, the dominant decay mode of the lightest scalar
Higgs boson is predicted to be $\hn\rta JJ$, where $J$ is the
(massless) Majoron.%
\Ref\majoron{J.C. Romao, F. de Campos, and J.W.F. Valle,
\plbj{292} (1992) 329.}\
$J$ interacts too weakly to be observed in the detector.
In such models, the decays of the second lightest scalar Higgs boson
can also be predominantly invisible, the two most important modes being
$JJ$ and $\hn\hn(\rta JJJJ)$.
\Ref\majoronii{J.C. Romao, J.L. Diaz-Cruz, F. de Campos, and J.W.F. Valle,
preprint FTUV-92-39 (1992).}

Two possible detection modes for an invisibly decaying Higgs boson
can be envisioned at a hadron collider.  Associated $Z\hn$ production
was considered in
\REF\fkane{S. Frederiksen, N. Johnson, G. Kane, and J. Reid,
SSCL-preprint-577 (1992), unpublished.}
Ref.~\fkane, with the rough conclusion that a viable
signal for $\hn\rta I$ ($I$ being any invisible channel) can be detected
for $\mhn\lsim 150\gev$ provided the $\hn ZZ$ coupling is Standard Model
(SM) strength and $BR(\hn\rta I)\sim 1$.
In this letter, we consider associated production of
top plus anti-top plus Higgs.  We find that if the top quark is not too light
($\mt\gsim 130\gev$) and if $BR(\hn\rta I)\sim 1$, then a viable signal for
$\hn\rta I$ can be extracted for $\mhn\lsim 250\gev$ when the $\hn t\anti t$
coupling is of SM strength. Clearly, the $Z\hn$ and $\tth$ modes are
complementary in the sense that they rely on the vector boson
vs. fermion couplings, respectively, of the Higgs boson.
For a CP-even Higgs boson,
which has both types of coupling, both modes tend to be viable,
but for a CP-odd Higgs boson the $ZZ$ coupling is
absent at tree-level and only the $\tth$ mode studied here
could lead to a visible signal.

Our procedure is quite simple.  We trigger on $t\anti t \hn$ events
by requiring that one of the $t$'s decay to a lepton
($e$ or $\mu$) with $\ptr>20\gev$ and $|\eta|<2.5$,
which is isolated from other jets and leptons by $\Delta R=0.3$.
In order to single out events containing a $t\anti t$ pair, we also demand
that at least one of the $b$-quarks be vertex tagged.
The efficiency and purity of $b$-tagging was studied by
the SDC collaboration.
\Ref\sdctdr{Solenoidal Detector Collaboration Technical Design Report,
E.L. Berger \etal, Report SDC-92-201, SSCL-SR-1215, 1992, p 4.15-4.16.}
Based on this study, we assume that
any $b$-jet with $|\eta|<2$ and $\ptr>30\gev$
will have a probability of 30\% (independent of $\ptr$) of being
vertex tagged, provided there is no other vertex within $\Delta R_V=0.5$.
(It may turn out that 30\% is too conservative a number; the SDC
TDR $b$-tagging efficiency can be increased by looking for
the lepton associated with a semi-leptonic $b$ decay.)
Misidentification backgrounds are not significant so long as
the probability to tag (\ie\ mis-identify)
a light quark or gluon jet as a $b$-jet
under these same conditions is of order 1\%, and the corresponding
number for $c$-quark jets is of order 5\%.
Jets with $\ptr$ below $30\gev$ are assumed not to be tagged.

To further single out the $t\anti t$ events of interest,
we demand that there be three or more jets
with $\ptr>30\gev$ and $|\eta|<2.5$ that are isolated from all other
jets by at least $\Delta R=0.7$.
\foot{For $b$'s that decay hadronically the jet momentum is taken
to be that of the $b$ quark; for $b$'s that decay semi-leptonically,
the jet momentum is computed by omitting the momentum carried
by the neutrino.}
(One of these jets is allowed to be the tagged $b$-jet.)
The invariant mass of each pair of
jets, $\mtwo$, is computed and at least one pair {\it not containing the
tagged $b$-quark} is required to have
$\mw-\delmw/2 \leq\mtwo\leq\mw+\delmw/2$. In addition,
we combine any pair of jets satisfying this criteria
with the tagged $b$ jet(s) and compute the three-jet invariant mass, $\mthree$.
We demand that $\mt-\delmt/2\leq\mthree\leq\mt+\delmt/2$ for at least
one $bjj$ combination.  Together, these two cuts greatly reduce
the likelihood that the second top in a $t\anti t$ event
can decay leptonically and satisfy all our criteria.
The results to be presented employ values of $\delmw=15\gev$ and
$\delmt=25\gev$. Only a small fraction of
signal events are eliminated by such mass cuts for typical
jet and lepton energy resolutions,
\foot{We employ $\delta E/E=0.5/\sqrt{E(GeV)}\oplus 0.03$ for jets
and $\delta E/E=0.2/\sqrt{E(GeV)}\oplus 0.01$ for leptons. The very first step
of our analysis is to smear the energies of all leptons and jets
using these resolutions.} whereas the reducible
backgrounds are significantly decreased.

Finally, to reveal the invisibly decaying Higgs, we determine the
missing transverse momentum, $\ptmiss$, for the event and compute
$\mmissl$, the transverse mass obtained by combining the
transverse components of the missing momentum and the lepton momentum,
$\mmissl^2\equiv(\etmiss+\etell)^2-(\ptmiss+\ptell)^2$.
The transverse missing momentum is computed by taking the incoming beam
momenta and subtracting from their sum the momenta of all observable
final state partons.  (Any jet falling outside $|\eta|=5$ is
deemed unobservable.) In addition, a probabilistically fluctuating
missing momentum from the underlying minimum bias event structure
is included. And, as already noted, jet energy and momenta are smeared.
The $t\anti t\hn$ events events of interest
are characterized by very broad distributions in $\mmissl$ and $\etmiss$.
Cuts on both variables will be made.

There are several sources of background.  The most obvious is the
irreducible background from $t\anti t Z$ events in which $Z\rta \nu\anti\nu$.
This background will be denoted by $\ttz$; its $\etmiss$ and $\mmissl$
distributions are very much like those of the signal.
The important reducible backgrounds all derive from various tails related
to $t\anti t$ or $t\anti t g$ events.   We shall artificially separate
the $t\anti t (g)$ backgrounds into two components. The first, and most
important, component is that
where both top quarks decay leptonically, in which case
only $t\anti t g$ events can possibly satisfy the
cuts outlined earlier.  We shall refer to this background as the $\ttglnu$
background. The $\tth$ signal, irreducible $\ttz$
background, and $\ttglnu$ background have in common the feature that
their $\mmissl$ and $\etmiss$ spectra
are essentially independent of whether or not the $b$ quarks
from the $t$'s decay semi-leptonically or purely hadronically.
In the second component, only one $t$ decays semi-leptonically.  However,
the $t\anti t(g)$ rate is so large that a not insignificant number of
events could survive our cuts simply due to the fact that
there is at least the one neutrino from the leptonically decaying
$W$ providing the trigger lepton, and perhaps additional neutrinos from
$b$ quarks that decay semi-leptonically.  The background from $t\anti t$ plus
$t\anti t g$ events, in which only one
$W$ coming from the $t$ quarks decays leptonically,
will be denoted by $\ttbdecay$.

The strategies required to control the $\ttglnu$ and $\ttbdecay$ backgrounds
are more or less `orthogonal'.
The $\ttglnu$ background is easily reduced to a level below that
of the $\ttz$ background by a cut on $\etmiss$.  Once such a cut is made,
however, the $\ttglnu$ background has a very long $\mmissl$ tail
and is not significantly reduced by cuts on this latter variable. Thus, the
optimal $\etmiss$ cut is essentially determined by the $\ttglnu$ background.
In the case of the $\ttbdecay$ background, if the only neutrino present
is that from the single leptonically decaying $W$, then there is a very sharp
Jacobian peak in $\mmissl$ near the $W$ mass. After including the off-shell
$W$ tail, this background still only populates $\mmissl$ values below
about 100 GeV.  This fact provides the main
motivation for employing a cut on $\mmissl$. Of course,
semi-leptonic $b$ decays lead to a tail to the $W$-decay Jacobian peak
in the $\mmissl$ spectrum. This tail can be significant
for $t\anti t g$ events (but is quite small for events in which
there is no extra radiated gluon), and
forces us to a somewhat higher cut on $\mmissl$.
However, it is easy to find an $\mmissl$ cut which reduces the
$\ttbdecay$ background to a negligible level while retaining
much of the $\tth$ signal (and $\ttz$ background).

Thus, it is straightforward to find cuts on $\etmiss$ and $\mmissl$ that are
adequate for uncovering the invisibly decaying Higgs. Nonetheless,
it may be useful to note that the $\ttglnu$ background could
be suppressed further by demanding that all events contain only one
identified {\it isolated} lepton.  Although $\tau$'s will be difficult to
identify, $\mu$'s can always be identified and $e$'s are easily identified
so long as they are not too soft and are isolated. Reduction
of the $\ttglnu$ background by a factor of
order 1/3 to 1/2 could probably be achieved, with essentially
no impact on the $\tth$ signal (or $\ttz$ background).
Note that it is probably
not advantageous to veto against non-isolated leptons (which can
probably be done only for $\mu$'s in any case).  Although such a veto
would tend to suppress the already small $\ttbdecay$ background somewhat,
it would also suppress the signal rate.
The results in this paper do not make use of any type of second-lepton veto.

Our computations will not include $t\anti t jj$ backgrounds ($j=g$ or $q$).
Such backgrounds are higher order in $\alpha_s$ and our normalization
procedure (described later)
is such that the $t\anti t +t\anti t g$ subprocesses should provide
a good leading order estimate of the backgrounds from $t\anti t $
events including radiated jet(s).
No new physical effects are introduced (for our cut procedures) by going
beyond one extra jet.

We also will not explicitly compute $t\anti t W$ or $t\anti b W$ backgrounds.
When the extra $W$ decays leptonically, such processes will yield
broad $\mmissl$ and $\etmiss$ spectra.
But these processes are higher order in the weak
coupling constant than the leading order $\ttglnu$ background discussed
earlier.
The $\ttz$ background can be significant since the $Z$ can decay
to automatically invisible $\nu\anti\nu$ final states, whereas $t\anti t W$
events are only a background if the explicit $W$ decays via
$W\rta \ell\nu$ and the $\nu$ carries most of the $W$ momentum.
In addition, the rate for $t\anti t W$ events
is proportional to $q^{\prime}\anti q$ luminosities, which are much smaller
at the LHC than the $gg$ luminosity responsible for $t\anti t Z$ production.
The $t\anti b W$ process can be thought of in part as $t\anti t$ production
in which an off-shell $t$ `decays' to $bW$. Clearly, this will be substantially
suppressed compared to the on-shell decay backgrounds that we examine.
In addition, the $\mthree$ cut is even more effective in eliminating
this background than in the case of the $\ttglnu$ and $\ttbdecay$
on-shell decay backgrounds.

A host of other backgrounds have been thought about and dismissed.
For example, $WWgg$ events are suppressed by being electroweak
in origin, by the need to mis-tag one of the $g$'s as a $b$,
and by a cut on $\mmissl$ (note that only one $W$ can decay leptonically
if we are to get at least three jets and if two non-tagged jets
are to have mass near $\mw$).

We have employed exact matrix element calculations for all the subprocesses.
All the production reactions we consider are dominated by $gg$ collisions.
We have employed distribution functions for the gluons evaluated at
a momentum transfer scale given by the subprocess energy.  It is
well-known that QCD corrections are substantial for $gg$ initiated processes.
For example, for the $gg\rta t\anti t$ process
the QCD correction `K' factor has been found to be of the order
of 1.6 for our choice of scale.
\Ref\qcdcor{P. Nason, S. Dawson, and R.K. Ellis,
\npbj{303}, 607 (1988).  See also, W. Beenakker \etal, \prdj{40},
54 (1989).}
Computations of the `K' factors for the other reactions we consider
are not yet available in the literature. We will assume that they
are of the same magnitude.  Our precise procedures follow.
Rates for the $t\anti t\hn$ and $t\anti t Z$
processes have been multiplied by a QCD correction factor of 1.6.
In the case of $t\anti t (g)$ we have incorporated
the `K' factor as follows. We have generated events without
an extra gluon ($t\anti t$ events) and have also generated events
with an extra gluon ($t\anti t g$ events) requiring that
the $\ptr$ of the extra gluon be $>30\gev$.  For this cutoff
one finds $\sigma(t\anti t g)\sim 0.6 \sigma(t\anti t)$.  Thus,
if the two event rates are added together without cuts
an effective `K' factor of 1.6 is generated.  As already described,
explicitly allowing
for an appropriate number of $t\anti t g$ events is important in
properly estimating the backgrounds to our $\mmissl$ distribution.
Our procedure should yield an upper limit
on the number of events with an extra gluon having $\ptr>30\gev$
and therefore potentially providing a background source
due to extra radiated jets in association with $t\anti t$ production.
The gluon distribution functions we have employed are the D0$^\prime$
distributions of
\REF\mrs{A.D. Martin, R.G. Roberts, and W.J. Stirling,
preprint RAL-92-078 (1992).} Ref.~\mrs.

\smallskip
\noindent{\bf 2. Results and Discussion}
\smallskip

\FIG\mtcompfig{}
\midinsert
\vbox{\phantom{0}\vskip 4.5in
\phantom{0}
\vskip .5in
\hskip +40pt
\special{ insert scr:hinvisible_mtcomp.ps}
\vskip -1.4in }
\centerline{\vbox{\hsize=12.4cm
\Tenpoint
\baselineskip=12pt
\noindent
Figure~\mtcompfig: Distribution shape in $\mthree$
for the $\mhn=100\gev$ signal (solid) compared to that for the
$\ttglnu$ background (dashes), at $\mt=140\gev$. We have taken
$\delmw=15\gev$, and required $\etmiss>200\gev$ and $\mmissl>250\gev$.
}}
\endinsert

In this letter, we focus on results for the LHC.
Some preliminary discussion is useful.
First, we note that
a plot of the $\mmissl$ event rate spectrum for the $\tth$ signal
as compared to that for the sum of the
$\ttz$, $\ttglnu$, and $\ttbdecay$ backgrounds
shows great similarity in shape
for the signal and background once $\mmissl$ is large enough that
the $\ttbdecay$ background is small. For $\mmissl\gsim 150\gev$,
signal and background both fall very slowly as $\mmissl$ increases.
This similarity implies the need for an accurate determination
of the expected background level.  The importance of a cut on $\mthree$
is illustrated in Fig.~\mtcompfig.  There,
we compare the $\mthree$ distribution shapes for the $\mhn=100\gev$
signal (the $\ttz$ background would yield a similar $\mthree$ plot)
and the (primary reducible) $\ttglnu$ background at $\mt=140\gev$;
in the figure we have required $\etmiss>200\gev$ and $\mmissl>250\gev$.
The $\tth$ signal exhibits a strong peak near $\mt$, whereas
the bulk of the $\ttglnu$ background populates much higher $\mthree$ values.

 \TABLE\lhclum{}
 \topinsert
 \titlestyle{\twelvepoint
 Table~\lhclum: Number of $100\fbi$ years (signal event rate)
 at LHC required for a $5\sigma$ confidence level signal, with cuts
specified by (see text) $\etmiss>200\gev$,
$\mmissl>150\gev$, $\delmw=15\gev$ and $\delmt=25\gev$.
$BR(\hn\rta I)=1$ is assumed.}
 \bigskip
 \hrule \vskip .04in \hrule
 \thicksize=0pt
 \begintable
  $\mt$\\$\mhn$ | 60 & 100 & 140 & 200 & 300 \cr
  110 |   1.3(  24) &   2.0(  31) &   2.8(  36) &   5.2(  49) &
  15.8(  86) \nr
  140 |   0.2(  18) &   0.4(  23) &   0.6(  30) &   1.5(  47) &
   4.3(  80) \nr
  180 |   0.2(  24) &   0.3(  31) &   0.5(  44) &   1.6(  76) &
   5.2( 139) \endtable
 \hrule \vskip .04in \hrule
 \endinsert

In order to quantify the observability of the Higgs signals,
we have estimated the number of LHC years required for a $\nsd=5$
sigma significance of the signal compared to background for $BR(\hn\rta I)=1$.
The statistical significance $\nsd$ is computed as $S/\sqrt B$.
For any given integrated luminosity, $S$ is the total $\tth$ event rate
and $B$ the total $\ttz+\ttglnu+\ttbdecay$
event rate, with $\etmiss>200\gev$ and $\mmissl>150\gev$
(and all other cuts) imposed. The required number of years as
a function of $\mhn$ for various $\mt$ values appears in Table~\lhclum.
Also given (in parentheses) is the associated number of signal events
($S$).  The associated number of background events ($B$)
can be obtained from the relation $B=S^2/25$.

{}From this table, it is immediately apparent that
detection of an invisibly decaying Higgs boson should be possible
within 1 to 2 LHC years for $\mt\gsim 130\gev$ and $\mhn\lsim 200-250\gev$
if (as assumed in these calculations)
its coupling to $t\anti t$ is of Standard Model strength.
(The required number of years for non-SM coupling is obtained
simply by dividing the results of Table~\lhclum\ by the ratio of the
$t\anti t$ coupling strength to the SM strength, raised to the 4th power.)
For $\mhn\gsim 300\gev$, the $\tth$ event rate drops to a lower level
such that more than 2 LHC years are required. However, it is rather
unlikely that invisible decays would be dominant for a Higgs boson
with mass above 200 GeV or so.

The $\mmissl$ and $\etmiss$ cuts chosen
for Table~\lhclum\ are probably near optimal.
If the $\mmissl$ cut is strengthened to $\mmissl>200\gev$, the $\ttbdecay$
background becomes completely negligible, but
the $\tth$ (and $\ttz$) event rates are cut by about 25\% while the $\ttglnu$
background falls only slightly; as a result, the time
required to achieve a 5 sigma effect typically increases by 15-25\%.
As noted earlier, weakening of the $\etmiss$ cut causes a significant
increase in the $\ttglnu$ background relative to signal.
Strengthening this latter cut to $\etmiss>250\gev$ improves
$S/B$, but at a sacrifice of signal which makes $\hn$ discovery statistically
more difficult.

Of course, the $\nsd$ values quoted assume that the normalization
of the expected background will be well-determined by the time
that the experiments are performed.  This will require a good
understanding of the missing energy tails as they actually appear
in the detectors, calculation of the higher-order QCD
corrections that we have only estimated,
and accurate knowledge of the parton (especially gluon)
distribution functions.  With the availability of HERA data, and through
the analysis and study of $t\anti t$ events in the actual detectors,
we believe that uncertainties in the relevant backgrounds can be brought down
to the 20\% level by the time that adequate luminosity has been
accumulated that an invisible Higgs signal would become apparent.

Remaining questions include the following. Will
the efficiency of $b$-tagging at the LHC, given the many overlapping
events expected, be as great as assumed here based on the SDC detector
study? Will the
missing-energy tails from overlapping events be significant? Will it
be possible to build
sufficiently hermetic detectors given radiation
damage issues, and so forth? Our results are sufficiently
encouraging that the LHC detector
collaborations should study carefully the impact of these issues upon this
detection mode.

Our studies have been performed for a $\hn$ that is a CP-even
Higgs mass eigenstate.
The $\tth$ rates would be somewhat different as a function of
$\mhn$ for a mixed CP or CP-odd eigenstate. However, we do not
anticipate that the results for such cases would differ by very
much from those obtained here.

\smallskip
\noindent{\bf 3. Conclusion}
\smallskip

We have demonstrated that at the LHC a hermetic detector
with the ability to tag $b$-quark jets
should allow detection of an invisibly decaying Higgs boson
with a $t\anti t\hn$ associated production rate comparable to
that of a SM Higgs boson. Allowing
for several years of running at canonical luminosity,
such detection should be possible for $\mhn\lsim 200-250\gev$.

\smallskip\noindent{\bf Acknowledgements}
\smallskip
This work has been supported in part by Department of Energy
grant \#DE-FG03-91ER40674
and by Texas National Research Laboratory grant \#RGFY93-330.
I am grateful to M. Barnett, H. Haber, S. Novaes and F. Paige
for helpful conversations. Some of the Monte Carlo generators employed were
developed in collaboration with J. Dai, L. Orr and R. Vega.
I would also like to thank
the Aspen Center for Physics, where a substantial portion of this work
was performed.

\smallskip
\refout
\end